\pdfoutput=1

\documentclass[letterpaper, 10 pt, conference]{ieeeconf}

\usepackage{graphicx}
\usepackage{booktabs}
\usepackage{adjustbox}
\usepackage{wrapfig}

\usepackage{subfigure}

\usepackage{lipsum}
\usepackage{amssymb}
\usepackage{amsmath}
\usepackage{siunitx}
\usepackage{mathtools}
\usepackage{xcolor}
\usepackage{hyperref}
\let\labelindent\relax
\usepackage[inline]{enumitem}
\usepackage[ruled,vlined,linesnumbered]{algorithm2e}
\usepackage[noend]{algpseudocode}
\usepackage{ifthen}
\usepackage{dsfont}
\usepackage[backend=biber,style=numeric-comp,sorting=none,maxbibnames=99]{biblatex}
\usepackage{balance}
\usepackage{dirtytalk}
\usepackage{afterpage}

\usepackage{times}

\allowdisplaybreaks

\newcommand{\E}{\mathbb{E}}

\newcommand{\Prob}{\mathbb{P}}

\newcommand{\ra}{\rightarrow}
\newcommand{\R}{\mathbb{R}}

\newtheorem{remark}{Remark}

\newtheorem{proposition}{Proposition}
\newtheorem{definition}{Definition}

\newtheorem{theorem}{Theorem}

\newtheorem{example}{Example}

\newcommand{\network}{\mathcal{N}}
\newcommand{\nodes}{I}

\newcommand{\arcs}{A}

\newcommand{\arcsLoadConstraintSet}{\mathcal{W}}
\newcommand{\arcLoad}{w}

\newcommand{\latency}{s}
\newcommand{\cost}{c}
\newcommand{\toll}{p}
\newcommand{\optTolls}{P}

\newcommand{\costToGo}{z}

\newcommand{\routeSetMod}{\textbf{R}}
\newcommand{\routes}{\routeSetMod}

\newcommand{\height}{h}

\newcommand{\nodeLoadIn}{g}

\usepackage{todonotes}
\usepackage{bbm}

\bibliography{references}

\IEEEoverridecommandlockouts
\title{\LARGE \bf
Approximately Optimal Toll Design for Efficiency and Equity in Arc-Based Traffic Assignment Models
}

\author{Chih-Yuan Chiu$^{1}$
\thanks{$^{1}$School of Electrical and Computer Engineering, Georgia Institute of Technology, GA 30332 (email: \texttt{cyc at gatech dot edu}).}
}

\begin{document}

\maketitle

\thispagestyle{empty}
\pagestyle{empty}

% \thispagestyle{plain}
% \pagestyle{plain}

% ~\\ \frank{Restore to pagestyle empty during final compile before submitting.} ~\\

%%%%%%%%%%%%%%%%%%%%%%%%%%%%%%%%%%%%%%%%%%%%%%%%%%%%%%%%%%%%%%%%%%%%%%%%%%%%%%%%

\begin{abstract}
Congestion pricing policies have emerged as promising traffic management tools to alleviate traffic congestion caused by travelers' selfish routing behaviors. The core principle behind deploying tolls is to impose monetary costs on frequently overcrowded routes, to incentivize self-interested travelers to select less easily congested routes. Recent literature has focused on toll design based on arc-based traffic assignment models (TAMs), which characterize commuters as traveling through a traffic network by successively selecting an outgoing arc from every intermediate node along their journey. However, existing tolling mechanisms predicated on arc-based TAMs often target the design of a single congestion-minimizing toll, ignoring crucial fairness considerations, such as the financial impact of high congestion fees on low-income travelers. To address these shortcomings, in this paper, we pose the dual considerations of efficiency and equity in traffic routing as bilevel optimization problems. Since such problems are in general computationally intractable to solve precisely, we construct a linear program approximation by introducing a polytope approximation for the set of all tolls that induce congestion-minimizing traffic flow patterns. Finally, we provide numerical results that validate our theoretical conclusions.
\end{abstract}

\section{Introduction and Related Work}
\label{sec: Introduction and Related Works}

Traffic congestion on urban highways has steadily worsened in recent years, leading to high economic costs 
\cite{LeBeau2019TrafficJamsUS} 
and giving rise to serious public health concerns \cite{CDPH2016TrafficExhaustPollutants}. Congestion pricing mechanisms have recently attracted increasing attention from state and federal transportation agencies as a promising traffic management tool that can alleviate rush-hour traffic \cite{Ley2023NYCCongestion, WinnieLey2024NewYorkTakesCrucialStep}. Such schemes impose a monetary toll on routes in the network that can easily become congested, in the hopes of persuading travelers to instead select alternative routes \cite{Roughgarden2002HowBadIsSelfishRouting}. In effect, congestion pricing schemes aim to alter the routing decisions of a population of selfish travelers, to induce traffic flow patterns in the network that correspond to lower levels of overall traffic congestion. 

The design of congestion pricing policies is often predicated upon a traffic assignment model (TAM), a mathematical framework that describes the manner in which self-interested travelers formulate routing decisions in a traffic network. Specifically, in route-based TAMs, travelers are modeled as selecting a single route from the set of all routes connecting their source and destination, without deviating from their selection mid-journey \cite{Roughgarden2002HowBadIsSelfishRouting, Stewart2007TollingTrafficLinksUnderStochasticAssignment, Maheshwari2024CongestionPricingforEfficiencyEquity}. Arc-based TAMs, on the other hand, regard travelers as sequentially choosing, at each intermediate node along their journey, the next arc (edge) that minimizes their travel cost-to-go among all available options. In recent years, arc-based TAMs have enjoyed rising popularity due to their greater empirical success at predicting travelers’ routing decisions, compared to conventional route-based routing models \cite{Oyama2022Markovian, AkamatsuOyama2023GlobalStability}. As such, recent literature has focused on toll design under arc-based traffic assignment models (TAMs). Specifically, \cite{Chiu2023DynamicTollinginArcBasedTAMs} designed tolls, based on an arc-based TAM, that would accurately capture gradual shifts in travelers’ routing preferences while provably inducing minimum congestion on the network.

However, existing congestion pricing schemes have also attracted criticism for imposing unduly heavy financial burdens on low-income travelers, due to their regressive nature \cite{Paybarah2019CongestionPricingWorkingClass}. Tolls imposed on freeway express lanes effectively form a financial barrier, preventing less wealthy users from accessing more convenient commuting options. Thus, when designing congestion management policies, a transit authority must balance the need to induce \textit{efficient} use of the traffic infrastructure, to minimize congestion, with the need to account for \textit{equity} considerations, to ensure that the deployed tolls do not pose an excessively heavy financial burden to low-income commuters. To achieve the dual objectives of efficiency and equity, in this work, we first characterize the set of all tolls that induce equilibrium flows that minimize congestion. Then we select, among these congestion-minimizing tolls, those that satisfy equity considerations, such as minimizing the maximum congestion fee that travelers would have to pay. 

Our work shares similarities with the methodology and analysis in \cite{Dial1999MinimalRevenueCongestionPricingPart1, Dial2000MinimalRevenueCongestionPricingPart2, Stewart2007TollingTrafficLinksUnderStochasticAssignment, Fleischer2004TollsForHeterogeneousSelfishUsers, Maheshwari2024CongestionPricingforEfficiencyEquity}, which present toll designs that account for both efficiency and equity considerations, but only for route-based TAMs and not for arc-based TAMs. Recent works have also focused on alleviating the inequity induced by current congestion pricing mechanisms by subsidizing low-income users’ access to tolled roads \cite{Chiu2024CreditvsDiscount}, or by establishing a rebate program \cite{Jalota2021WhenEfficiencyMeetsEquity} that invests toll revenues in public transit projects or refunds toll revenues to low-income users. In contrast, in this work, we directly optimize the congestion fees, to ensure they induce congestion-minimizing traffic flows while minimizing their impact on low-income commuters.

% are no more than necessary to induce congestion-minimizing traffic flows.

The rest of this paper is structured as follows. 
% Sec. \ref{Related Works} surveys recently proposed methods for equitable congestion pricing and situates this work in the relevant literature. 
Sec. \ref{sec: Preliminaries} presents the arc-based TAM, cost structure and equilibrium traffic flow concepts considered throughout this work. Sec. \ref{sec: Toll Design for Efficiency and Equity} frames the dual objectives of promoting the efficient and equitable use of traffic infrastructure as a bilevel optimization problem, where the outer-level optimization task accounts for equity considerations, and the inner-level optimization task accounts for efficiency objectives. To approximately solve such bilevel optimization problems in a computationally tractable manner, in Sec. \ref{sec: Polytope Approximation for Congestion-Minimizing Tolls}, we introduce a polytope approximation of the set of congestion-minimizing tolls, which reduces the bilevel optimization problem to a linear program. Finally, in Sec. \ref{sec: Experiments}, we validate our theoretical results on simulated traffic networks, and in Sec. \ref{sec: Conclusion and Future Work}, we summarize our results and present promising future work directions.

\section{Preliminaries}
\label{sec: Preliminaries}

This section presents the traffic network structure (Sec. \ref{subsec: Network Structure}) and traveler utility models considered in this work.

\subsection{Network Structure}
\label{subsec: Network Structure}

Consider a finite acyclic directed graph $\network = (\nodes, \arcs)$ representing a single-origin, single-destination traffic network, where $\nodes$ and $\arcs$ respectively denote the set of nodes and arcs (i.e., edges) in the network $\network$. Let $o \in \nodes$ and $d \in \nodes$ respectively represent the origin and destination nodes of $\network$, and let $g_o \geq 0$ denote the traveler flow into the traffic network at the origin node $o$. Given any node $i \in \nodes$, we respectively represent the sets of incoming and outgoing arcs at $i$ by $\arcs_i^- \subseteq \arcs$ and $\arcs_i^+ \subseteq \arcs$. Travelers journey through the network by successively selecting a sequence of arcs from the origin $o$ to the destination $d$. 

In subsequent sections, physical quantities such as the cost incurred by travelers on each arc will be defined recursively throughout the network, often from the destination $d$ to the origin $o$ using the principle of dynamic programming. These recursive definitions require the following concept of the \textit{height} of an arc in a directed acyclic traffic network, as introduced in \cite[Def. 1]{Chiu2023ArcbasedTrafficAssignment}. 
For each $r \in \routes$ and arc $a \in r$, let $\height_{a,r}$ denote the location of arc $a$ in the route $r$ when listing arcs in the route $r$ backwards from the destination. In other words, from origin to destination, $a$ is the $(|r| - \height_{a,r})$-th arc in route $r$. With a slight abuse of notation, we denote by $\height_a := \max_{r \in \routes: a \in r} \height_{a,r}$ the \textit{height} of arc $a$. In words, $\height_a$ is the length of the longest route segment connecting $i_a$, the start node of $a$, to the destination node $d$. We denote by $\height(\network)$ the length of the longest route in the network $\network$.

% For each arc $a \in \arcs$, let $\routes_a$ denote the set of all route segments $\{a_1, \cdots, a_k\}$ that start at arc $a$ and end at the destination node $d$, i.e., $a_1 = a$, $j_{a_k} = d$, and $j_{a_s} = i_{a_{s+1}}$ for each $s \in \{1, \cdots, k-1\}$. For each arc $a \in \arcs$, we define the \textit{height} of arc $a \in \arcs$, denoted $\height_a$, to be the length of the longest route segment in $\routes_a$. Let $\routes$ denote the set of all routes connecting the origin $o$ and the destination $d$ in the network $\network$, i.e., $\routes = \routes_a$ for any $a \in \arcs$ satisfying $i_a = o$. We denote by $\height(\network)$ the length of the longest route in the network.

The heights of arcs in a network $\network$ satisfy the following properties, first introduced in \cite[Prop. A.2]{Chiu2023ArcbasedTrafficAssignment}.

\begin{proposition}(\cite[Prop. A.2]{Chiu2023ArcbasedTrafficAssignment}, Restated) \label{Prop: Condensed DAG Properties, Height}
Given a directed acyclic traffic network $\network = (\nodes, \arcs)$ with the route set $\routes$:
\begin{enumerate}
    \item For any $a \in \arcs$, we have $\height_a = 1$ if and only if $j_a = d$. Similarly, if $\height_a =\height(\network)$, then $i_a = o$.
    
    \item For any $r \in \routes$, and any $a, a' \in r$ with $\height_{a, r} < \height_{a',r}$, we have $\height_a < \height_{a'}$ i.e., arcs along a route from the origin to the destination have strictly decreasing depth.
    
    \item Fix any $a \in \arcs$, and any $r \in \routes$ containing $a$ such that $\height_{a,r} = \height_a$. Then, for any $a' \in \routes$ following $a$ in $r$, we have $\height_{a',r} = \height_{a'}$.
    
    \item For each height $k \in [\height(\network)] := \{1, \cdots, \height(\network)\}$, there exists an arc $a \in \arcs$ such that $\height_a = k$.
\end{enumerate}
\end{proposition}

\begin{proof}
See Chiu et al. \cite{Chiu2023ArcbasedTrafficAssignment}, Section A.2.
\end{proof}

\subsection{Costs and Equilibria}
\label{subsec: Costs and Equilibria}

In the traffic network $\network = (\nodes, \arcs)$, each arc $a \in \arcs$ is characterized by a latency cost $\latency_a: \R \ra \R$ incurred by each traveler unit, which describes the cost (to each traveler unit) of the time required to traverse through the arc $a$ as a function of the traffic flow level (load) $\arcLoad_a$ on arc $a$. We assume that each latency cost function $\latency_a(\cdot)$ is strictly positive, strictly increasing, and strictly convex, to capture the real-life observation that higher traffic densities induce longer commute times. 

To reshape traveler incentives and induce more socially desirable traffic flow patterns, the central traffic authority deploys a set of (non-negative) tolls $\toll \in \R_{\geq 0}^{|\arcs|}$ in the network, where, for each arc $a \in \arcs$, $\toll_a \geq 0$ denotes the toll value assigned to each arc $a \in \arcs$. Then, for each arc $a \in \arcs$, the total \textit{travel cost} of $\cost_a: \R^2 \ra \R$ incurred by all the travelers on arc $a \in \arcs$ can be obtained by summing the latency and toll costs:
\begin{align*}
    \cost_a(\arcLoad_a, \toll_a) &:= \latency_a(\arcLoad_a) + \toll_a.
\end{align*}
However, travelers may in general lack access to real-time, perfect knowledge of traffic conditions throughout the entire network. To account for variability in travelers' perception of their latency costs, we append a zero-mean noise term $\delta_a$ to the travel cost $\cost_a$, to form the notion of the \textit{perceived} total travel cost $\tilde \cost_a: \R^2 \ra \R$, as defined below:
\begin{align*}
    \tilde c_a(\arcLoad_a, \toll_a) &:= \latency_a(\arcLoad_a) + \toll_a + \delta_a.
\end{align*}
At each node $i \in \nodes$ except the destination node, travelers choose their next arc from the set of all outgoing arcs $a \in \arcs_i^+$ at node $i$. In particular, the probability of selecting each outgoing arc $a \in \arcs_i^+$ depends on the perceived minimum cost-to-go $\tilde z_a$ of that arc, recursively defined backwards the destination node $d$:
\begin{align*}
    \tilde \costToGo(\arcLoad, \toll)
    = \begin{cases}
        \tilde \cost_a(\arcLoad_a, \toll_a), &j_a = d, \\
        \tilde \cost_a(\arcLoad_a, \toll_a) + \E_\delta \Big[ \min\limits_{a' \in \arcs_{j_a}^+} \tilde \costToGo_a(\arcLoad, \toll) \Big], &j_a \ne d.
    \end{cases}
\end{align*}
In this work, we use the logit Markovian noise model, which draws the noise terms $\delta_a$ from the Gumbel distribution with scale (entropy) parameter $\beta > 0$. In this case, the expected cost-to-go $z_a := \E_\delta[\tilde z_a]$ for each arc $a \in \arcs$ is computed as:
\begin{align*}
    \costToGo_a(\arcLoad, \toll) &= \cost_a(\arcLoad, \toll) - \frac{1}{\beta} \ln\Bigg( \sum_{a' \in \arcs_{j_a}^+} e^{-\beta z_{a'}(\arcLoad, \toll)} \Bigg),
\end{align*}
and the corresponding probability with which a traveler at each non-destination node $i \in \nodes \backslash \{d\}$ selects from the outgoing arcs $a \in \arcs_i^+$ is given as follows. For each $a \in \arcs_i^+$:
\begin{align*}
    \Prob\Big(z_a = \min\limits_{a' \in \arcs_i^+} z_{a'} \Big) &= \frac{e^{-\beta z_a(\arcLoad, \toll)}}{\sum_{a' \in \arcs_i^+} e^{-\beta z_{a'}(\arcLoad, \toll)}}.
\end{align*}
At steady state, the traveler population's routing decisions in aggregate generate an equilibrium flow pattern known as the \textit{Markovian traffic equilibrium}, first introduced in Baillon and Cominetti \cite{BaillonCominetti2008MarkovianTrafficEquilibrium}.

We first introduce the traffic flow constraint set $\arcsLoadConstraintSet$, which enforces the continuity of traffic flows at each non-destination node $i \in \nodes$, by:
\begin{align} \label{Eqn: Def, W}
    \arcsLoadConstraintSet := &\Bigg\{ \arcLoad \in \R^{|\arcs|}: \sum_{a \in \arcs_i^+} \arcLoad_a = \sum_{a \in \arcs_i^-} \arcLoad_a, \hspace{0.5mm} \forall \hspace{0.5mm} i \ne o, d, \\ \nonumber
    &\hspace{1cm} \sum_{a \in \arcs_o^+} \arcLoad_a = \nodeLoadIn_o, \hspace{1mm} \arcLoad_a \geq 0, \hspace{0.5mm} \forall \hspace{0.5mm} a \in \arcs \Bigg\}.
\end{align}
Given any set of fixed tolls $\toll \in \R_{\geq 0}^{|\arcs|}$, the corresponding \textit{Markovian Traffic Equilibrium (MTE)} $\bar \arcLoad^\beta(\toll)$ is defined to be the unique flow vector $\arcLoad \in \arcsLoadConstraintSet$ that satisfies the fixed-point equation below. At any non-destination node $i \in \nodes \backslash \{d\}$, and for any outgoing arc $a \in \arcs_i^+$:
\begin{align*}
    \arcLoad_a &= \left(g_i + \sum_{a' \in \arcs_{i}^+} \arcLoad_{a'} \right) \cdot  \frac{\exp(-\beta z_a(\arcLoad, \toll))}{\sum_{a' \in \arcs_i^+} \exp(-\beta z_{a'}(\arcLoad, \toll))}.
\end{align*}
Above, $g_i = g_o$ if $i = o$; otherwise, $g_i = 0$.

\section{Toll Design for Efficiency and Equity}
\label{sec: Toll Design for Efficiency and Equity}

In this section, we first characterize the space of \textit{congestion-minimizing tolls}, denoted $\optTolls$, which contains all toll values that induce equilibrium traffic flow patterns which minimize the overall network congestion (Sec. \ref{subsec: Congestion-Minimizing Tolls}). We then present a bilevel optimization problem that encodes our search for a congestion-minimizing toll $\toll \in \optTolls$ which optimizes a pre-determined fairness objective (Sec. \ref{subsec: Bilevel Optimization for Efficient, Equitable Toll Design}).

\subsection{Congestion-Minimizing Tolls}
\label{subsec: Congestion-Minimizing Tolls}

The objective of designing and deploying tolling policies is to induce a corresponding equilibrium flow pattern that minimizes overall network congestion in the following sense. 

\begin{definition}(\textbf{Congestion-Minimizing Toll}) \label{Def: Congestion-Minimizing Toll}
Define the perturbed total latency of the network by $L: \arcsLoadConstraintSet \ra \R$, where:
\begin{align} \label{Eqn: Collective, Total Latency}
    L(\arcLoad) &:= \sum_{a \in \arcs} \arcLoad_a \latency_a(\arcLoad_a) \\ \nonumber
    &\ + \frac{1}{\beta} \Bigg[ \sum_{a \in \arcs_i^+} \arcLoad_a \ln \arcLoad_a - \Bigg(\sum_{a \in \arcs_i^+} \arcLoad_a \Bigg) \ln \Bigg(\sum_{a \in \arcs_i^+} \arcLoad_a \Bigg)\Bigg].
\end{align}
We call $\arcLoad^\star := \text{arg}\min_{\arcLoad \in \arcsLoadConstraintSet} L(\arcLoad)$ the congestion-minimizing flow vector, and we call $\toll \in \R_{\geq 0}^{|\arcs|}$ a congestion-minimizing toll if $\bar \arcLoad^\beta(\toll) = \text{arg}\min_{\arcLoad \in \arcsLoadConstraintSet} L(\arcLoad) = \arcLoad^\star$.
\end{definition}

In the following discussion, let $\optTolls$ denote the set of congestion-minimizing tolls, i.e.,:
\begin{align} \label{Eqn: Opt Toll Set}
    \optTolls := \big\{\toll \in \R^{|\arcs|}: \bar \arcLoad^\beta(\toll) = \text{arg} \min_\arcLoad L(\arcLoad) \big\}
\end{align}

\begin{remark}
The first summand in the definition of $L(w)$ (Def. \ref{Def: Congestion-Minimizing Toll}) is the weighted sum of all arc latencies, with weights given by arc flows, and thus describes the collective latency experienced by the entire traveler population throughout the network. The second term is a regularization term that captures the noisy optimality of the travelers' decision-making. This regularization term is included to model the traffic flow pattern that would result if the population of noisily optimal travelers were to instead cooperate to minimize the total network latency, rather than selfishly selecting routes to minimize their overall travel costs. Since latency functions are strictly positive, strictly increasing, and strictly convex by assumption, and the regularization term is strictly convex as well \cite{Chiu2023ArcbasedTrafficAssignment, Chiu2023DynamicTollinginArcBasedTAMs}, $L(\cdot)$ is strictly convex. 
% For details, see \cite{Chiu2023ArcbasedTrafficAssignment, Chiu2023DynamicTollinginArcBasedTAMs}.
\end{remark}

\subsection{Bilevel Optimization for Efficient, Equitable Toll Design}
\label{subsec: Bilevel Optimization for Efficient, Equitable Toll Design}

% ~\\
% \frank{Exposition: ~\
% \begin{enumerate}
%     \item Deploying the congestion-minimizing marginal toll $\toll^\star$ achieves the efficiency objective of minimizing overall congestion, but ignores other metrics of societal welfare such as fairness or equity.
%     \item In particular, the marginal toll may impose undue financial burdens on the traveler population, especially on lower-income travelers (maybe add a small example here?)
%     \item Other tolls in $\optTolls$ may induce the same minimum congestion without exacerbating societal inequity
%     \item To alleviate congestion on traffic networks without exacerbating societal inequities, 
%     \item 
% \end{enumerate}
% }
% ~\\
The design of the congestion-minimizing marginal toll $\toll^\star$ in Sec. \ref{subsec: Congestion-Minimizing Tolls} achieves the efficiency objective of reducing overall congestion in the traffic network, but ignores other metrics of societal welfare such as fairness or equity. In particular, the marginal toll may assume unnecessarily large values on network edges on which traffic flows are already efficiently routed, as the following example indicates.

\begin{example}
For a parallel network with edges whose latency functions are identical, the MTE equilibrium involves a uniform allocation of input flow across edges, which also corresponds to the congestion-minimizing flow pattern. Thus, no toll needs to be deployed to induce the socially optimal outcome. On the other hand, the marginal tolls $p^\star$ corresponding to such networks are in general strictly positive. Thus, although deploying $p^\star$ would likewise induce MTE equilibrium flows that are congestion-minimizing, such tolls pose an undue financial burden on low-income travelers. 
\end{example}

In general, while designing tolls to alleviate congestion, the traffic authority should also take into account equity considerations in addition to efficiency metrics. Below, we present a list of reasonable metrics $\{F(\arcLoad, \toll): i = 1, 2, 3\}$ for societal welfare that pertain to traffic routing.
\begin{enumerate}
    \item \textbf{Minimum-Revenue (Min-Rev)}---Minimize the total toll revenue that the entire traveler population expends on their journey:
    \begin{align} \label{Eqn: F1, Min Rev}
        F_1(\arcLoad, \toll) &:= \sum_{a \in \arcs} \arcLoad_a \toll_a.
    \end{align}
    The minimization of $F_1$ above ensures that the deployed tolls are 
    % the minimum congestion fees
    no more than
    necessary to induce latency-minimizing traffic flows. $F_1$ is analogous to the minimum-revenue objective explored in Dial et al. \cite{Dial1999MinimalRevenueCongestionPricingPart1, Dial2000MinimalRevenueCongestionPricingPart2} and Stewart et al. \cite{Stewart2007TollingTrafficLinksUnderStochasticAssignment}, albeit modified to fit the arc-based TAM formulation considered in this work. 

    \item \textbf{Minimum Max-Cost (Min-Max)}---Minimize the least upper bound for the total (i.e., toll plus latency) travel cost that any traveler may incur along their journey:
    \begin{align} \label{Eqn: F2, Min Max}
        F_2(\arcLoad, \toll) &:= \max_{r \in \routes} \sum_{a \in r} \big[ \latency_a(\arcLoad_a) + \toll_a \big].
    \end{align}
    Above, $F_2$ computes the worst-case total latency and toll costs incurred by the traveler population along any route connecting the origin and destination nodes.
    
    % \item \textbf{Minimum Cost Gap  (Min-Gap)}---Minimize the gap between the maximum and minimum possible total (i.e., toll plus latency) travel cost that any traveler may incur along their journey:
    % \begin{align} \label{Eqn: F3, Min Gap}
    %     &F_3(\arcLoad, \toll) \\ \nonumber
    %     := \ &\max_{r, r' \in \routes} \Bigg| \sum_{a \in r} \big[ \latency_a(\arcLoad_a) + \toll_a \big] - \sum_{a' \in r'} \big[ \latency_{a'}(\arcLoad_{a'}) + \toll_{a'} \big] \Bigg|.
    % \end{align}
    % The above \say{Min-Gap} objective is analogous to the equity metric considered in Maheshwari et al. \cite{Maheshwari2024CongestionPricingforEfficiencyEquity}, albeit adapted to fit the arc-based TAM setting considered in this work.

    \item \textbf{Minimum Relative Entropy of Tolls over Routes}---Minimize the relative entropy of the tolls incurred over feasible routes, as given by:
    {
    \small
    \begin{align} \label{Eqn: F3, Min Relative Entropy of Tolls}
        F_3(\arcLoad, \toll) &:= 
        \sum_{r \in \routes} \sum_{a \in r} \toll_a \log \left( \frac{\sum_{a \in r} \toll_a}{\sum_{r' \in \routes} \sum_{a \in r'} \toll_{a'}} \right) + \Vert \toll \Vert_2^2.
        % \max_{r \in \routes} \sum_{a \in r} \big[ \latency_a(\arcLoad_a) + \toll_a \big].
    \end{align}
    }
    
    Minimizing $F_3$ would compel the total toll $\sum_{a \in r} \toll_a$, incurred by a traveler along a selected route $r \in \routes$, to assume similar values across routes $r \in \routes$. Although minimizing $F_3$ would unfortunately also promote high toll values\footnote{Note that the logarithmic term in \eqref{Eqn: F3, Min Relative Entropy of Tolls} is always non-positive.}, we control toll magnitudes by appending the regularization term $\Vert \toll \Vert_2^2$.
    
\end{enumerate}

Given a particular fairness metric $F: \arcsLoadConstraintSet \ra \R$, the traffic authority aims to design a toll $\toll \in \R^{|\arcs|}$ that minimizes $F: \arcsLoadConstraintSet \ra \R$, \textit{while also minimizing the overall latency} $L(\arcLoad)$, as defined by \eqref{Eqn: Collective, Total Latency} in Def. \ref{Def: Congestion-Minimizing Toll}. In other words, the traffic authority faces the following bilevel optimization problem:
\begin{align} \label{Eqn: Bilevel opt}
    \min_{\toll \in \R_{\geq 0}^{|\arcs|}} \hspace{3mm} &F\big( \bar \arcLoad^\beta(\toll), \toll \big) \\ \nonumber
    \text{s.t.} \hspace{3mm} &\bar \arcLoad^\beta(\toll) = \text{arg} \min_{\arcLoad} L(\arcLoad) = \arcLoad^\star.
\end{align}

The subsequent section (Sec. \ref{sec: Polytope Approximation for Congestion-Minimizing Tolls}) presents methods for approximately solving \eqref{Eqn: Bilevel opt}.

\section{Polytope Approximation for Congestion-Minimizing Tolls}
\label{sec: Polytope Approximation for Congestion-Minimizing Tolls}

Below, we present a tractable approximation to the bilevel optimization problem of designing tolls to achieve both efficiency and equity objectives in congestion management, as presented in \eqref{Eqn: Bilevel opt}. First, in Sec. \ref{subsec: Polytope Subset of Congestion-Minimizing Tolls}, we introduce a polytope approximation $\tilde \optTolls$ of the set $\optTolls$ of congestion-minimizing tolls. Then, in Sec. \ref{subsec: LP Approximations to the Bilevel Optimization Problem}, we approximate the bilevel optimization problem \eqref{Eqn: Bilevel opt} of achieving both efficiency and equity objectives in traffic routing by replacing $\optTolls$ with $\tilde \optTolls$. We then prove that the resulting approximate optimization problem (see \eqref{Eqn: Bilevel opt approx, with tilde P}) can be cast as a linear program under certain choices of the societal cost objective $F$.

\subsection{Polytope Subset of Congestion-Minimizing Tolls}
\label{subsec: Polytope Subset of Congestion-Minimizing Tolls}

First, we rewrite the bilevel optimization problem \eqref{Eqn: Bilevel opt} encoding the traffic authority's efficiency and equity aims as:
\begin{align} \label{Eqn: Bilevel opt, with P}
    \min_{\toll \in \optTolls} \hspace{5mm} &F\big( \arcLoad^\star, \toll \big),
\end{align}
where $\optTolls$ denotes the set of congestion-minimizing tolls (Def. \ref{Def: Congestion-Minimizing Toll}). Note that $\optTolls$ can be difficult to characterize explicitly, since it is implicitly defined by the optimization problem of minimizing the cost $L$. Below, we first assert that $P$ is non-empty, and then derive a polytope approximation $P_s$ of $P$ that allows \eqref{Eqn: Bilevel opt, with P} to be tractably solved.

The following result, established by \cite{Chiu2023ArcbasedTrafficAssignment, Chiu2023DynamicTollinginArcBasedTAMs}, shows that $\optTolls$ is non-empty, i.e., there exists at least one congestion-minimizing toll vector. In particular, the specific toll introduced below parallels the well-studied notion of the \textit{marginal toll} in the literature, albeit refined to accommodate the MTE equilibrium concept under consideration in this paper.

\begin{proposition}[\textbf{\cite{Chiu2023DynamicTollinginArcBasedTAMs}, Theorem 2}] \label{Prop: Congestion-Minimizing Marginal Toll}
There exists a toll $\toll^\star \in \R^{|\arcs|}$ that uniquely satisfies the fixed-point equation given by:
\begin{align} \label{Eqn: Congestion-Minimizing Marginal Toll}
    \toll^\star &= \bar \arcLoad^\beta(\toll^\star) \cdot \frac{d \latency_a}{d \arcLoad_a} \big( \bar \arcLoad^\beta(\toll^\star) \big).
\end{align}
Moreover, $\toll^\star \in \optTolls$. We call $\toll^\star$ the \textit{MTE marginal toll}.
\end{proposition}

% The following proposition 
Below, we 
show that $\optTolls$ can be approximated by a ($|\nodes|-1$)-dimensional polytope passing through $\toll^\star$.

\begin{theorem} \label{Thm: Polytope Approximation of Congestion-Minimizing Tolls}
Fix $\tau_i \in \R$ arbitrarily for each $i \in \nodes \backslash \{d\}$, and set $\tau_d := 0$. 
Define:
\begin{align} \label{Eqn: Opt Toll Set, Approximate}
    \tilde \optTolls &:= \{ (p_a^\star + \tau_{i_a} - \tau_{j_a})_{a \in \arcs}: \tau \in \R^{|\nodes|}, \tau_d = 0 \} \bigcap \R_{\geq 0}^{|\arcs|}.
\end{align}
Then $\tilde \optTolls \subseteq P$.
\end{theorem}

\begin{proof}
Let $\toll \in \tilde \optTolls$, i.e., $\toll_a \geq 0$ for each $a \in \arcs$, and there exists some $\tau \in \R^{|\nodes|}$ such that $\tau_d = 0$ and $\toll_a = \toll_a^\star + \tau_{i_a} - \tau_{j_a}$ for each $a \in \arcs$. We first claim that, for each $a \in \arcs$, we have $\costToGo_a(\arcLoad^\star, \toll) = \costToGo_a(\arcLoad^\star, \toll^\star) + \tau_{i_a}.$ Indeed, for any arc $a \in \arcs$ of height 1, we have:
\begin{align} \nonumber
    \costToGo_a(\arcLoad^\star, \toll) &= \latency_a(\arcLoad_a^\star) + \toll_a = \latency_a(\arcLoad_a^\star) + \toll_a^\star + \tau_{i_a} \\ \label{Eqn: z relation, 1}
    &= \costToGo_a(\arcLoad^\star, \toll^\star) + \tau_{i_a}.
\end{align}
Now, suppose $\costToGo_a(\arcLoad^\star, \toll) = \costToGo_a(\arcLoad^\star, \toll^\star) + \tau_{i_a}$ for any arc $a \in \arcs$ of height $h$ or lower, for some $h \in \{1, \cdots, \height(\network)-1\}$. Then, for any arc $a \in \arcs$ of height $h+1$:
\begin{align} \label{Eqn: z relation, 2}
    \costToGo_a(\arcLoad^\star, \toll) &= \latency_a(\arcLoad_a^\star) + \toll_a - \frac{1}{\beta} \ln \Bigg( \sum_{a' \in \arcs_{j_a}^+} e^{-\beta \costToGo_{a'}(\arcLoad^\star, \toll)} \Bigg) \\ \label{Eqn: z relation, 3}
    &= \latency_a(\arcLoad_a^\star) + \toll_a^\star + \tau_{i_a} - \tau_{j_a} \\ \nonumber
    &\hspace{1cm} - \frac{1}{\beta} \ln \Bigg( \sum_{a' \in \arcs_{j_a}^+} e^{-\beta \big[ \costToGo_{a'}(\arcLoad^\star, \toll) + \tau_{i_{a'}} \big]} \Bigg) \\ \label{Eqn: z relation, 4}
    &= \latency_a(\arcLoad^\star) + \toll_a^\star + \tau_{i_a} - \tau_{j_a} \\ \nonumber
    &\hspace{1cm} - \frac{1}{\beta} \ln \Bigg( e^{-\beta \tau_{j_a}} \cdot \sum_{a' \in \arcs_{j_a}^+} e^{-\beta \costToGo_{a'}(\arcLoad^\star, \toll)} \Bigg) \\ \nonumber
    &= \costToGo_a(\arcLoad^\star, \toll^\star) + \tau_{i_a},
\end{align}
where \eqref{Eqn: z relation, 3} follows from \eqref{Eqn: z relation, 2} by substituting $\toll_a = \toll_a^\star + \tau_{i_a} - \tau_{j_a}$ and \eqref{Eqn: z relation, 1}, while \eqref{Eqn: z relation, 4} follows from \eqref{Eqn: z relation, 3} by observing that for any arc $a \in \arcs$ and $a' \in \arcs_{j_a}^+$, we have $i_{a'} = j_a$.

Next, we claim that $\bar \arcLoad^\beta(\toll) = \arcLoad^\star$, and thus $\toll \in \tilde \optTolls$. By definition of the MTE, since $\bar \arcLoad^\beta(\toll^\star) = \arcLoad^\star$, we have:
\begin{align} \label{Eqn: z relation, 5}
    \frac{\arcLoad_a^\star}{\sum_{a' \in \arcs_{i_a}^+} \arcLoad_{a'}^\star} &= \frac{e^{-\beta \costToGo_a(\arcLoad^\star, \toll^\star)}}{\sum_{a' \in \arcs_{i_a}^+} e^{-\beta \costToGo_{a'}(\arcLoad^\star, \toll^\star)}} \\ \label{Eqn: z relation, 6}
    &= \frac{e^{-\beta \big[\costToGo_a(\arcLoad^\star, \toll^\star) + \tau_{i_a} \big]}}{\sum_{a' \in \arcs_{i_a}^+} e^{-\beta \big[\costToGo_{a'}(\arcLoad^\star, \toll^\star) + \tau_{i_{a'}} \big] }} \\ \label{Eqn: z relation, 7}
    &= \frac{e^{-\beta \costToGo_a(\arcLoad^\star, \toll)}}{\sum_{a' \in \arcs_{i_a}^+} e^{-\beta \costToGo_{a'}(\arcLoad^\star, \toll)}},
\end{align}
where \eqref{Eqn: z relation, 5} follows from the definition of the MTE, while \eqref{Eqn: z relation, 7} follows from \eqref{Eqn: z relation, 6} since, for any fixed $a \in \arcs$ and any $a' \in \arcs_{i_a}^+$, we have $i_{a'} = i_a$. Thus, $\arcLoad^\star$ satisfies the fixed-point equation that (uniquely) characterizes the MTE corresponding to $\toll$, so $\arcLoad^\beta(\toll) = \arcLoad^\star$.
\end{proof}

In words, Theorem \ref{Thm: Polytope Approximation of Congestion-Minimizing Tolls} states that the set of congestion-minimizing tolls $P$, which may in general be difficult to characterize, is provably (under-)approximated by the polytope $\tilde \optTolls$, obtained by intersecting the non-negative orthant $\R_{\geq 0}^{|\arcs|}$ with an affine subspace passing through the marginal toll $\toll^\star$. In Sec. \ref{subsec: LP Approximations to the Bilevel Optimization Problem} below, we show that by replacing $\optTolls$ with $\tilde \optTolls$ in the reformulated bilevel problem \eqref{Eqn: Bilevel opt, with P}, we can approximately solve \eqref{Eqn: Bilevel opt} and \eqref{Eqn: Bilevel opt, with P} in a computationally tractable manner.

\subsection{LP Approximations to the Bilevel Optimization Problem}
\label{subsec: LP Approximations to the Bilevel Optimization Problem}

Here, we consider the approximation to the reformulated bilevel optimization problem \eqref{Eqn: Bilevel opt, with P} obtained by replacing $\optTolls$ with $\tilde \optTolls$, as defined below:
\begin{align} \label{Eqn: Bilevel opt approx, with tilde P} 
    \min_{\toll \in \tilde \optTolls} \hspace{3mm} F(\arcLoad^\star, \toll).
\end{align}
Note that since $\tilde \optTolls \subseteq \optTolls$, the minimum value of the optimization problem \eqref{Eqn: Bilevel opt approx, with tilde P} yields an upper bound for the minimum value of the optimization problem \eqref{Eqn: Bilevel opt, with P}.

We now consider the specific setting in which the cost function under consideration, $F(\arcLoad, \toll)$, is composed of a linear combination of the minimum-revenue objective $F_1(\arcLoad, \toll)$ given by \eqref{Eqn: F1, Min Rev}, the minimum max-cost objective $F_2(\arcLoad, \toll)$ given by \eqref{Eqn: F2, Min Max}, and the $L_2$-regularized minimum relative toll entropy cost $F_3(\arcLoad, \toll)$. Concretely, suppose $F(\arcLoad, \toll)$ takes the form:
\begin{align} 
% \nonumber
\label{Eqn: F lambda, Composite Objective}
    F_\lambda(\arcLoad^\star, \toll) &:= \sum_{k=1}^3 \lambda_k F_k(\arcLoad^\star, \toll),
\end{align}
with $\lambda_1, \lambda_2, \lambda_3 \geq 0$. Then, using straightforward optimization techniques (see \cite{BoydVandenberghe2004ConvexOptimization}, Chapter 4), we can rewrite \eqref{Eqn: Bilevel opt approx, with tilde P} as:

% We remark that when $F$ is the linear combination of the objectives $F_1$ (Minimum-Revenue), $F_2$ (Minimum Max-Cost), and $F_3$ (Minimum Relative Toll Entropy), as defined in \eqref{Eqn: F1, Min Rev}, \eqref{Eqn: F2, Min Max}, and \eqref{Eqn: F3, Min Relative Entropy of Tolls}, respectively, \eqref{Eqn: Bilevel opt approx, with tilde P} can be recast as a convex program. 
\looseness=-1
Concretely, if we set $F(\arcLoad^\star, \toll) := F_\lambda(\arcLoad^\star, \toll) = \sum_{k=1}^3 \lambda_k F_k(\arcLoad^\star, \toll)$ for some $\lambda_1, \lambda_2, \lambda_3 \geq 0$, \eqref{Eqn: Bilevel opt approx, with tilde P} can be written as:
\begin{alignat}{2} \label{Eqn: Bilevel opt, F lambda} 
    &\hspace{6mm} \min_{\toll \in \tilde \optTolls} \hspace{1cm} &&F_\lambda(\arcLoad^\star, \toll) \\ \nonumber
    = \ &\min_{(\toll, \tau) \in \R^{|\arcs| + |\nodes|}} &&\tilde F_\lambda(\toll, t) \\ \nonumber
    & \hspace{1cm} \text{s.t.} &&\sum_{a \in r} \big[ s_a(\arcLoad_a^\star) + \toll_a \big] \leq t, \\ \nonumber
    & &&\toll_a = \toll_a^\star + \tau_{i_a} - \tau_{j_a}, \ \forall \ a \in \arcs, \\ \nonumber
    & && \toll_a \geq 0, \ \forall \ a \in \arcs, \\ \nonumber
    & && \tau_d = 0,
\end{alignat}
where:
\begin{align} \nonumber
    &\tilde F_\lambda(\toll, t) \\ \nonumber
    := \ &\lambda_1 \cdot \sum_{a \in \arcs} \arcLoad_a^\star \toll_a + \lambda_2 t \\ \nonumber
    &\hspace{3mm} + \lambda_3 \left( \sum_{r \in \routes} \sum_{a \in r} \toll_a \log\left( \frac{\sum_{a \in r} \toll_a}{\sum_{r' \in \routes} \sum_{a \in r'} \toll_{a'} } \right) + \Vert \toll \Vert_2^2  \right)
\end{align}
In words, when $F = F_\lambda$ is a linear combination of $F_1$, $F_2$, and $F_3$ with non-negative coefficients $\lambda_1$, $\lambda_2$, and $\lambda_3$, respectively, the bilevel optimization problem \eqref{Eqn: Bilevel opt} can be posed as a convex program. Moreover, if $\lambda_3 = 0$, \eqref{Eqn: Bilevel opt} can be posed as a linear program.

\section{Experiments}
\label{sec: Experiments}

This section presents empirical results which demonstrate that, for a given traffic network, the linear program approximation \eqref{Eqn: Bilevel opt approx, with tilde P} to the bilevel optimization problem \eqref{Eqn: Bilevel opt} can be used to design tolling strategies that are effective at promoting the efficient use of traffic infrastructure without worsening societal inequities. In particular, the tolls computed by \eqref{Eqn: Bilevel opt approx, with tilde P} induce the same minimum level of overall congestion in a given traffic network, while imposing less financial strain on the traveler population compared to the marginal toll (Def. \ref{Def: Congestion-Minimizing Toll}).

\begin{figure}[ht]
     \centering
    \subfigure
    []
    {\includegraphics[
     width=0.22\textwidth
     ]{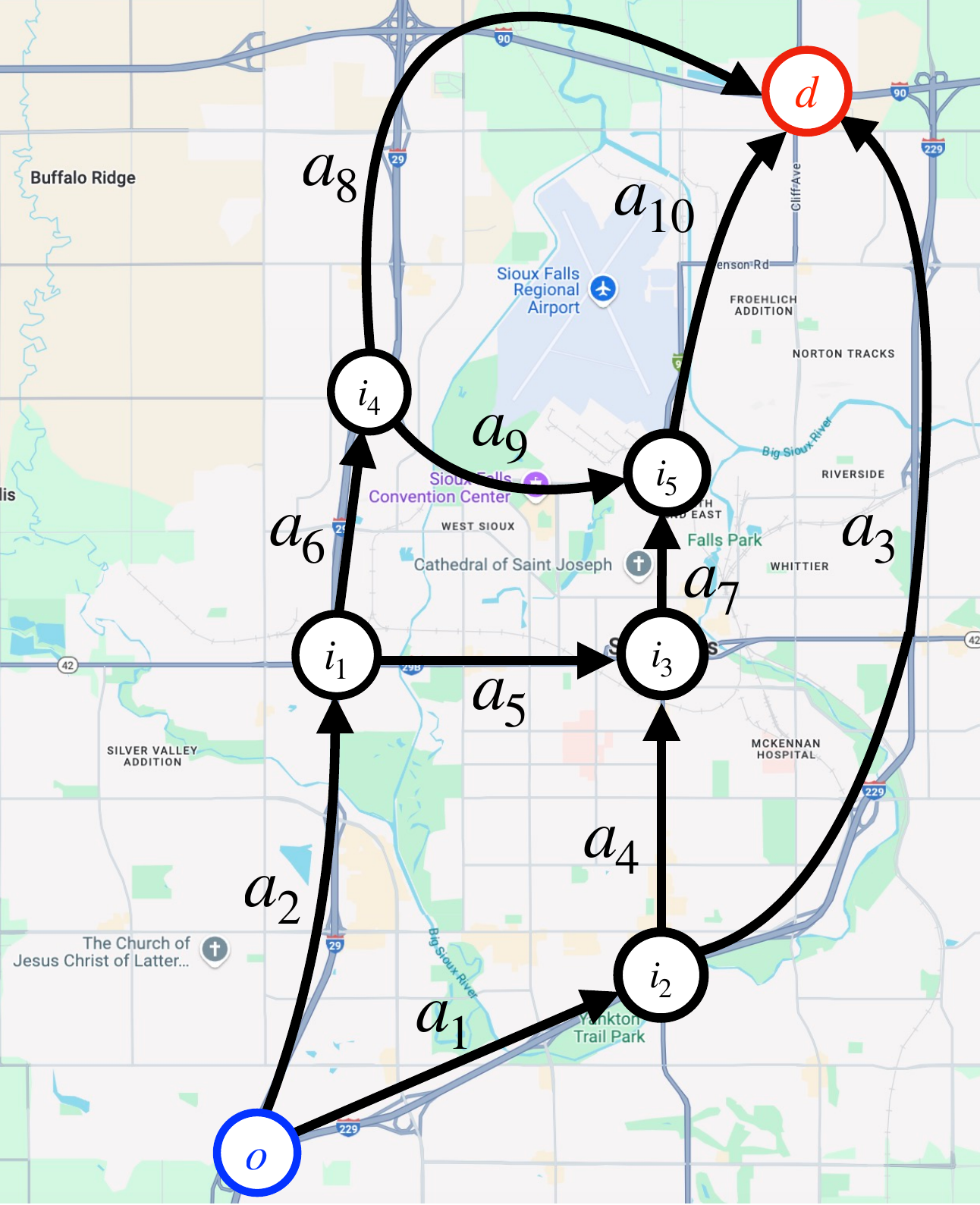}
     \label{fig:Sioux_Falls}
     }
    \subfigure
    []
    {\includegraphics[
    width=0.2\textwidth
    ]{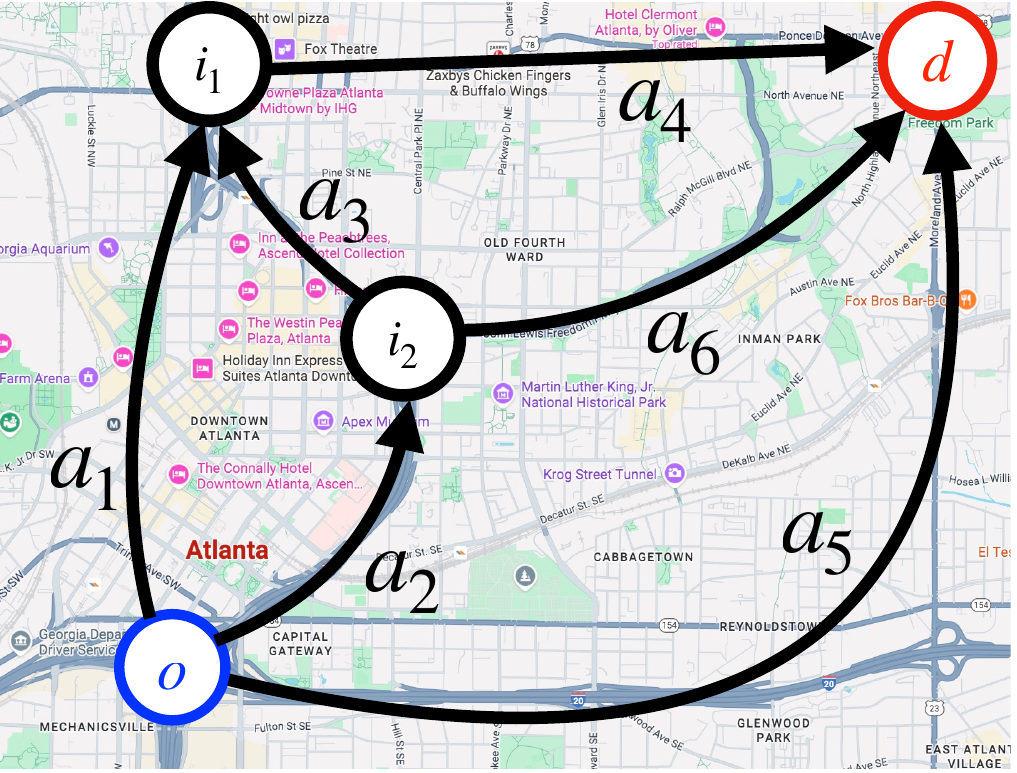}
    \label{fig:Atlanta_downtown}
    }
    \caption{Schematics for highway or freeway routes connecting an origin-destination pair in
    (a) Sioux Falls and
    (b) downtown Atlanta.
    }
    \label{fig:Sioux_Falls_Atlanta_downtown_Schematics}
\end{figure}

Specifically, consider the two network examples presented in Figs. \ref{fig:Sioux_Falls} and \ref{fig:Atlanta_downtown}, which represent highway and freeway routes connecting an origin-destination pair in the downtown areas of Sioux Falls, South Dakota, and Atlanta, Georgia, respectively. For both networks, we model the latency functions $\ell_a: \R \ra \R$ for each arc $a$ using the linear form $\ell_a(\arcLoad) = \theta_{a,1} \arcLoad_a + \theta_{a,0}$, where the parameters $\theta_{a, 1}$ and $\theta_{a, 0}$, across arcs $a$, are as listed in Table \ref{table:Latency_function_parameters}. For each network, we set the input flow at $g_o = 10$ and the entropy parameter at $\beta = 0.5$. As in Sec. \ref{subsec: LP Approximations to the Bilevel Optimization Problem}, we consider a composite cost function $F_\lambda(\arcLoad^\star, \toll) := \sum_{k=1}^3 \lambda_k F_k(\arcLoad^\star, \toll)$ with non-negative coefficients $\lambda_1, \lambda_2, \lambda_3 \geq 0$. For each network in Figs. \ref{fig:Sioux_Falls} and \ref{fig:Atlanta_downtown}, we then computed the marginal tolls $\toll^\star \in \R^6$ corresponding to the network, as well as the tolls derived by solving the following optimization problem:
\begin{align} \label{Eqn: Bilevel opt approx, with F lambda, tilde P}
    \min_{\toll \in \tilde \optTolls} \hspace{3mm} F_\lambda (\arcLoad^\star, \toll).
\end{align}
We then observe that, across a wide range of weights $\lambda \in \R^3$, the toll obtained by solving \eqref{Eqn: Bilevel opt approx, with F lambda, tilde P} induces user equilibrium flows corresponding to much lower values of the minimum-revenue ($F_1$) and minimum max-cost ($F_2$) objectives (Table \ref{table:MinRev_MinMax_Results}), compared to the marginal toll $\toll^\star$. In other words, the tolls computed by solving \eqref{Eqn: Bilevel opt approx, with tilde P} outperform the marginal tolls in minimizing the financial burden they would pose to the traveler population, while inducing the same congestion-minimizing traffic flow pattern that the marginal tolls would have induced.

\setlength{\tabcolsep}{4pt}
\begin{table}[t]
\centering
\caption{
Latency function parameters for the networks in Figs. \ref{fig:Sioux_Falls} and \ref{fig:Atlanta_downtown}.
% , respectively. 
}
% \scriptsize
% \vspace{-5pt}
\begin{tabular}{c|cc|cc}
\toprule%\hline
        \multicolumn{1}{c|}{} & 
        \multicolumn{2}{c|}{Network in Fig. \ref{fig:Sioux_Falls}} & 
        \multicolumn{2}{c}{Network in Fig. \ref{fig:Atlanta_downtown}} \\
        Arc ($a$) & $\theta_{a,1}$ & $\theta_{a,0}$ & $\theta_{a,1}$ & $\theta_{a,0}$ \\
        \midrule
        1 & 7.26 & 0.014 & 2.0 & 4.0 \\
        2 & 7.24 & 0.008 & 1.0 & 2.0 \\ 
        3 & 11.00 & 0.013 & 0.5 & 2.0 \\
        4 & 10.80 & 0.005 & 1.5 & 2.0 \\ 
        5 & 3.68 & 0.006 & 2.5 & 6.0 \\
        6 & 7.17 & 0.011 & 3.0 & 4.0 \\ 
        7 & 3.66 & 0.010 & (N/A) & (N/A) \\ 
        8 & 14.34 & 0.009 & (N/A) & (N/A) \\ 
        9 & 10.88 & 0.008 & (N/A) & (N/A) \\ 
        10 & 14.37 & 0.011 & (N/A) & (N/A) \\ 
        \bottomrule
    \end{tabular} \label{table:Latency_function_parameters}
\vspace{-10pt}
\end{table}

\setlength{\tabcolsep}{4pt}
\begin{table}[t]
\centering
\caption{
% \footnotesize 
% \frank{To edit; current table values and caption are incorrect.}
Minimum revenue ($F_1$) and miminum max-cost ($F_2$) objective values corresponding to the marginal toll ($\toll^\star$) and to the tolls minimizing the objective $F_\lambda(\arcLoad^\star, \toll) = \sum_{k=1}^3 \lambda_k F_k(\arcLoad^\star, \toll)$, for the networks illustrated in Fig. \ref{fig:Sioux_Falls} and \ref{fig:Atlanta_downtown}.
}
% \scriptsize
% \vspace{-5pt}
% \begin{tabular}{c|l|ccc}
\begin{tabular}{c|l|cc}
\toprule
%\hline
        % \multicolumn{1}{c|}{Weights} & \multicolumn{2}{c|}{Optimal CBCP} & \multicolumn{3}{c|}{\% using express lane} & \multicolumn{2}{c}{Average TT}  \\
        Network & \hspace{9mm} Toll $(\toll)$ & $F_1(\arcLoad^\star, \toll)$ & $F_2(\arcLoad^\star, \toll)$ 
        % & $\Vert \toll \Vert_1$ 
        \\
        \midrule
        Fig. \ref{fig:Sioux_Falls} & $\toll = \toll^\star$ & 101.68 & 27.21 
        % & 31.32
        \\
        & $\lambda = (1, 0, 0)$ & 28.88 & 20.66  
        % & 11.93 
        \\
        & $\lambda = (0, 1, 0)$ & 28.88 & 20.66 
        % & 11.93 
        \\
        & $\lambda = (0.7, 0, 0.3)$ & 28.90 & 20.66 
        % & 11.93 
        \\
        & $\lambda = (0, 0.7, 0.3)$ & 37.60 & 21.09 
        % & 11.93 
        \\
        & $\lambda = (0.5, 0.3, 0.2)$ & 28.88 &  20.66
        % & 11.93 
        \\
        Fig. \ref{fig:Atlanta_downtown} & $\toll = \toll^\star$ & 904.09 & 187.56
        % & 9.90
        \\
        & $\lambda = (1, 0, 0)$ & 273.82 & 129.67 
        % & 0.97 
        \\
        & $\lambda = (0, 1, 0)$ & 361.46 & 142.91 
        % & 0.97 
        \\
        & $\lambda = (0.7, 0, 0.3)$ & 274.44 & 129.71
        % & 0.97 
        \\
        & $\lambda = (0, 0.7, 0.3)$ & 275.14 & 129.71
        % & 0.97 
        \\
        & $\lambda = (0.5, 0.3, 0.2)$ & 273.79 & 129.67 
        % & 0.97 
        \\
        \bottomrule
    \end{tabular} \label{table:MinRev_MinMax_Results}
\vspace{-10pt}
\end{table}

\section{Conclusion and Future Work}
\label{sec: Conclusion and Future Work}

This work presents a computationally tractable method to solve the bilevel optimization problem of designing congestion pricing schemes that can minimize traffic congestion throughout a transportation network, while promoting equitable access to the traffic infrastructure.

There are many promising directions of future research. First, it is important to characterize the degree to which the polytope $\tilde \optTolls$ under-approximates the true set of congestion-minimizing tolls $\optTolls$. Second, since it is unrealistic to impose congestion fees on every arc in a real-life traffic network, a practical direction of future work would be to design \textit{second-best} congestion pricing mechanisms, i.e., toll-based policies in which only a subset of arcs can be tolled, that can provably induce efficient and equitable traffic flow configurations. Finally, it would be of interest to extend the approximation scheme and analysis in Sections \ref{sec: Toll Design for Efficiency and Equity} and \ref{sec: Polytope Approximation for Congestion-Minimizing Tolls} to encompass multi-origin, multi-destination, and cyclic networks.

% ~\\
% \frank{Future Work Directions:
% \begin{enumerate}
%     \item Multi-origin, multi-destination, cyclic network.
%     \item Exploring how much this subspace under-approximates the set of all latency-minimizing tolls.
%     \item Dynamics.
%     \item Second-best vs. First-best tolls.
%     \item Other metrics - Revenue maximization.
%     \item Large-scale experiments.
% \end{enumerate}
% }
% ~\\

\section{Acknowledgements} The author thanks Dr. Chinmay Maheshwari, Pan-Yang Su, and Dr. Shankar Sastry for fruitful discussions regarding the arc-based congestion game model studied in this work.

\printbibliography

\end{document}